\documentstyle[prl,aps,multicol,psfig,epsfig]{revtex}

\begin{document}
\draft
\author{V.\ M.\ Pudalov$^{a}$, M.\ E.\ Gershenson$^{b}$, H.\ Kojima$^{b}$,
G.\ Brunthaler$^c$, and G.\ Bauer$^c$}
\address{
$^{a}$ P.\ N.\ Lebedev Physics Institute, 119991 Moscow, Russia \\
$^{b}$ Serin Physics Lab, Rutgers University, Piscataway NJ-08854, USA\\
$^c$ Johannes Kepler Universit\"{a}t, Linz, A-4040, Austria}

\title{Reply to Comment
(cond-mat/0311174)} \maketitle

\begin{abstract}
We demonstrate that the experimental data on the temperature
dependence of the resistivity $\rho(T)$ and conductivity
$\sigma(T)$ for high-mobility Si-MOS structures, obtained for a
wide range of densities ($\sim 1:7$) at intermediate temperatures,
agree quantitatively with  theory of interaction corrections in
the ballistic regime, $T\tau >1$. Our comparison does not involve
any fitting parameters.

\end{abstract}
\vspace{-0.05in}
\begin{multicols}{2}
Recently \cite{aleiner}, we compared the resistivity $\rho(T)$
data for high mobility Si MOSFETs with the interaction corrections
calculated to the first order in $T$ \cite{ZNA}, and found {\em a
quantitative agreement} over a wide range of densities  at
intermediate temperatures (in the ballistic regime $T\tau >
\hbar/k_B$) \cite{aleiner,ZNA}. In the comparison, we used
independently determined values of the renormalized $g$-factor and
effective mass \cite{gm}.\\

In the  Comment by Shashkin, Dolgopolov, and Kravchenko (hereafter
SDK) \cite{SDKcomment}, several claims have been made; we respond
to the claims below:

(1) SDK claim that our paper \cite{aleiner} contradicts the
earlier publication \cite{weakloc}. In fact, Ref.~\cite{aleiner}
focuses on electron transport in the ballistic regime ($T\tau >
\hbar/k_B$), whereas Ref.~\cite{weakloc} considers the
single-particle interference in the diffusive regime
($\tau_\varphi \gg\tau$);  it has been shown that the weak
localization and interaction corrections in the diffusive regime
are irrelevant to the ``metallic'' $\rho(T)$. This conclusion does
not contradict our analysis in Ref.~\cite{aleiner}.

(2) SDK are concerned about our comparison with the theory of
interaction corrections by Zala, Narozhny, and Aleiner (hereafter
ZNA) \cite{ZNA} in terms of $\Delta\rho(T)$, and claim that
``there is no linear-in $T$ interval'' in our $\sigma(T)$ data.

Wherever the $T$-dependent corrections to the conductivity are
small ($\Delta\sigma(T)/\sigma \ll 1$), one can treat
$\Delta\rho(T)$ and $\Delta\sigma(T)$ on equal footing. At
sufficiently large densities ($n>3.5\times 10^{11}$ cm$^{-2}$),
this condition is fulfilled over a wide interval of temperatures
($\hbar/2\pi \tau < T_{\rm low}, \quad T_{\rm high} \ll T_F, \quad
T_{\rm high}-T_{\rm low} \sim$ a few K); this enables to detect
unambiguously the linear-in-$T$ regime of $\sigma(T)$. In
Ref.~\cite{aleiner} we analyzed resistivities, whereas here we
compare the data for both samples from Ref.~\cite{aleiner} with
the theory in terms of $\sigma(T)$. Figures~1,\,2 show that the
$\sigma(T)$ data are in a quantitative agreement with the
interaction corrections to the conductivity
$\sigma=\sigma_D+\Delta\sigma(T)$ (solid lines). In the
calculations, we have used the values for the density-dependent
renormalized effective mass $m^*/m_b$ and $g$-factor $g^*/g_b =
(\chi^*/\chi^b) (m_b/m^*)$, determined in independent experiments
\cite{gm}; the value of $\tau$ for each curve is obtained from the
experimental $\sigma_D\equiv (\rho(T\rightarrow 0))^{-1}$ data via
the extrapolation procedure described in Refs.~\cite{aleiner,ZNA}.
Once the $\tau$ value (and, hence, the Drude conductivity
$\sigma_D$) is fixed,
 the slope $d\sigma/dT$ is obtained with no free parameters.

\begin{figure}
\vspace{-0.2in}
 \centerline{\psfig{figure=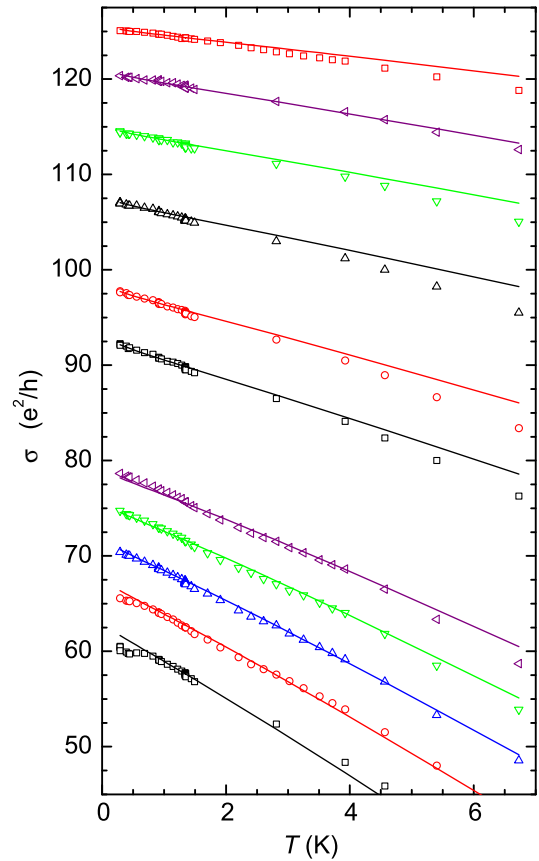,width=240pt}}
\vspace{-0.1in}
\begin{minipage}{3.2in}
\caption{Comparison of the $\sigma(T)$ data (symbols) for sample
Si22 reproduced from Ref.~\protect\cite{aleiner} with calculated
quantum corrections \protect\cite{ZNA}
(solid
lines).
Densities, from top to bottom, are (in units of $10^{11}$cm$^{-2}$):
21.3, 18.9, 16.5, 14.1, 11.7, 10.5, 8.1, 7.5, 6.9, 6.3, 5.7.}
\end{minipage}
\label{Fig.1}
\end{figure}

In the comparison, in accord with Ref.~\cite{das03}, we considered
only the temperature range $\hbar/2\pi\tau < T \ll T_F$, and
ignored the lower-$T$ and higher-$T$ data. As was discussed in
Ref.~\cite{aleiner}, the temperature range where the agreement
holds, shrinks towards low densities. This might be expected
because, as the density is decreased, the $\rho(T)$ dependence
becomes so steep that the temperature changes of
$\Delta\rho/\rho_D$ remain small only within a narrow range of
$T$. The agreement of the data with the interaction corrections in
Figs.~1,\,2 holds over a reasonable range of temperatures and in a
wide range of densities (1:7), but not too close to the critical
density value $n_c\sim 10^{11}$cm$^{-2}$.

\begin{figure}
\vspace{-0.1in}
 \centerline{\psfig{figure=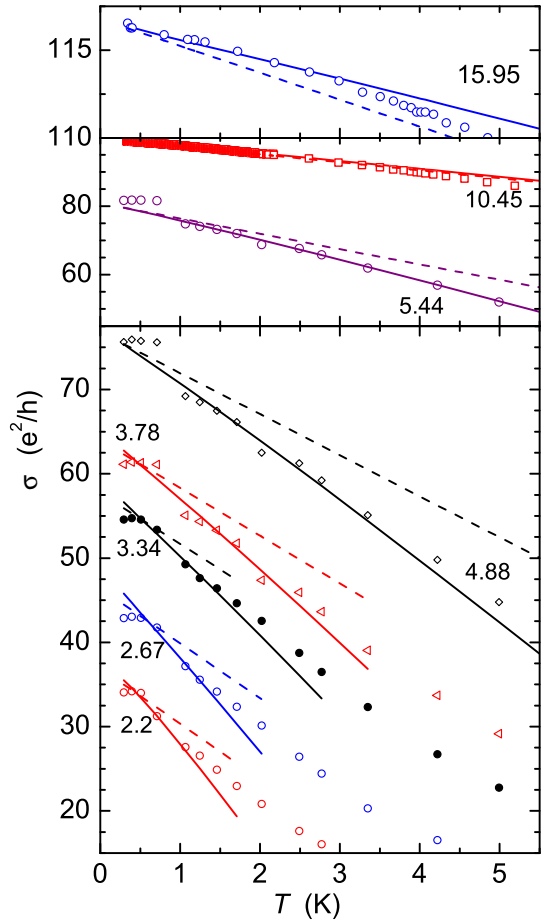,width=220pt}}
\vspace{0.1in}
\begin{minipage}{3.2in}
\caption{Comparison of the $\sigma(T)$ data (symbols) for sample
Si15 \protect\cite{aleiner} with the interaction corrections
\protect\cite{ZNA} (solid lines) and the ``screening'' theory
~\protect\cite{GD} (dashed lines). Densities are indicated in
units of $10^{11}$cm$^{-2}$.}
\end{minipage}
\label{Fig.2}
\end{figure}

Once again, no matter how good (or bad) the agreement at low
densities is (this regime is beyond the applicability of the
theory~\cite{ZNA} anyway), the main conclusion of our paper
\cite{aleiner} remains the same: the ``metallic'' behavior of
$\rho(T)$ in high-mobility Si MOSFETs in the regime $\rho \ll
h/e^2$ is in quantitative agreement with the
predictions~\cite{ZNA} based on considering the interaction
effects in the ballistic regime.
\vspace{0.1in}

(3) SDK claim that in the range of densities
$n>2\times10^{11}$cm$^{-2}$, the $\rho(T)$ changes can be
described by the ``traditional screening'' theory by one of the
authors \cite{GD}.

The ``screening'' theory \cite{GD} predicts the dependence
\begin{equation}
\delta\sigma(T)/\sigma(0)=1-2 C(\alpha)C(n)(T/E_F)+...,
\end{equation}
 with
$C(\alpha) =2\ln 2$ (or $(8/3) \ln 2$) for charged impurities,
$\alpha=-1$ (or for surface roughness, $\alpha=0$) scattering,
respectively, $C(n)=
F(2k_F)(1-G(2K_F)/(F(2k_F)[1-G(2k_F)]+2k_F/q_s)$, and $G(2k
_F)=0.2236$. By re-writing $T$ in units $x=T\tau /\hbar$
\cite{ZNA}, and $\sigma$ in units of $e^2/(\pi\hbar)$, we obtain
\begin{equation}
d\sigma/dx  \approx - 2 C(\alpha,n)=-4C(\alpha)C(n).
\end{equation}
This expression differs from that derived in Ref.~\cite{ZNA}:
\begin{equation}
d\sigma/dx \approx
15\frac{F_0^a}{1+F_0^a}\left[1-\frac{3}{8}t(x)\right] +
\left[1-\frac{3}{8}f(x)\right],
\end{equation}
where the functions $f(x)$ and $t(x)$ are defined in
Ref.~\cite{ZNA}.

The difference between the predictions of the ``screening'' theory
\cite{GD} and the theory of interaction corrections \cite{ZNA} is
illustrated in Figs.~2,3. Note that the difference in the slopes
$d\sigma/dT$ in Fig.~2 is a factor of $\sim 2$ at the lowest
density $n=2.2\times 10^{11}$cm$^{-2}$ (to calculate the
interaction corrections to $\sigma$, we have used the experimental
values of $F_0^a(n)$ and $m^*(n)$ measured by us in independent
experiments \cite{gm}). The theoretical predictions for the slopes
$d\sigma/dx$ as functions of the density are shown in Fig.~3. The
two density dependences cross each other at different $n$
depending on the specifics of impurity or interface scattering in
the screening theory. Adjusting the latter parameters does not
improve agreement between the theory \cite{GD} and our
experimental data, if a sufficiently wide range of densities is
considered. At the same time, the ZNA theory \cite{ZNA} is
consistent with the data in the wide density range without any
adjustable parameters. Theoretical aspects of the difference
between approaches \cite{GD} and \cite{ZNA} have been clarified
in Ref.~\cite{ZNA}.

\begin{figure}
\vspace{-0.05in}
\centerline{\psfig{figure=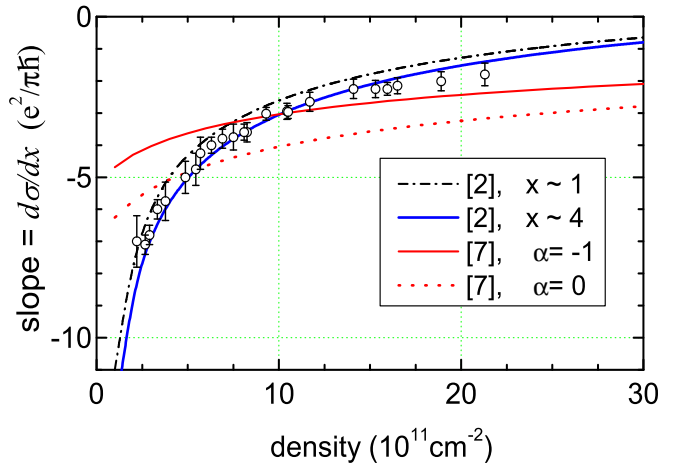,width=260pt,height=190pt}}
\vspace{0.1in}
\begin{minipage}{3.2in}
\caption{Comparison of the slopes $d\sigma/dx$, calculated from
Ref.~\protect\cite{GD} for two models of scatterers ($\alpha=-1$
and $\alpha=0$), and from Ref.~\protect\cite{ZNA} with $F_0^a$ and
$m^*$ values determined in \protect\cite{gm}. In the latter case,
the slope slightly varies with the temperature (compare $x=1$ and
$x=4$), due to the
functions $f(x)$  and $t(x)$ in
Eq.~(2). Symbols depict the slope $d\sigma/dx$  of the data shown
in Figs.~1,\,2.}
\end{minipage}
\label{Fig.3}
\end{figure}

(4) SDK suggest that the low-temperature deviation of our
$\rho(T)$ data from the theory, observed for several samples, was
caused by electron overheating. Apparently, they overlooked the
corresponding discussion in our paper (p.3, left column
\cite{aleiner}), which analyzed and rejected this possibility. For
the range of relatively high densities $n\gg n_c$ we explored in
Ref.~\cite{aleiner}, the deviations from the theory occur at high
temperatures, where overheating is highly unlikely. Moreover, the
low temperature part of the data at $T \lesssim \hbar/2\pi\tau$ is
irrelevant to the  linear-in-$T$ dependence \cite{das03}.
\vspace{0.1in}

(5) SDK claim that the range of densities explored in
\cite{aleiner} is ``not related to the anomalous increase of
$\rho(T)$ with temperature''.  The term ``metallic'' refers to the
sign of $d\rho/dT$, not to its magnitude; however, the ``anomalous
metallic'' behavior is more commonly referred to the strong
changes of $\rho(T)$ \cite{das03}. We emphasize that, though we
have limited our quantitative analysis in \cite{aleiner} to the
small changes $\Delta\rho(T)/\rho \ll 1$, where the theory
\cite{ZNA} is applicable, the overall increase of the resistance
with $T$ is large in the whole explored range of densities: $\rho$
increases with temperature by a factor of 2 even at $n$ as high as
$15\times10^{11}$cm$^{-2}$ \cite{weakloc}.

Note that this claim by SDK is inconsistent with their own
publications: indeed, in Ref.~\cite{SKD_r(T)}, they fitted the
linear $\sigma(T)$ dependences at low densities and low
temperatures ($T<1$K) with the theory~\cite{ZNA}, whereas on pages
2,3 of Ref.~\cite{kk}, one of the authors concluded on the basis
of analysis of $\sigma(T)$ over the \emph{same} ranges of
densities and temperatures that ``the linear $\rho(T)$ dependence
is different from … the dependence seen at higher temperatures and
higher densities''. \vspace{0.1in}

(6) SDK claim that the theory \cite{ZNA} ``is inapplicable at high
density $n>5\times 10^{11}$cm$^{-2}$, where the surface roughness
scattering becomes dominant''. In fact, the theory \cite{ZNA}
assumes the short-range scattering  (for more details, see
Ref.~\cite{Gornyi}). This requirement is well justified for our
samples over the entire range of $n=(2.2-30)\times 10^{11}$
cm$^{-2}$ as it has been shown in Ref.\cite{aleiner}: the ratio of
the ``large-angle'' scattering time $\tau$ to the quantum life
time (or ``all-angle'' scattering time) $\tau_q=\hbar/(2\pi T_D)$
is close to unity  (here, $T_D$ is the Dingle temperature).
Besides, it is expected that the difference between the effects of
short-range and long-range disorder on $\rho(T)$ is negligible at
$T< 0.1T_F$ ~\cite{das03}.

(7) Criticizing our analysis of $\rho(T)$ at ``high'' densities
$n\gg n_c$,  SDK give preference to comparison with the theory at
low densities $n\approx n_c$ and low temperatures. An attempt to
analyze $\rho(T)$ in this regime on the basis of the theory
\cite{ZNA}, undertaken by SDK in Ref.~\cite{SKD_r(T)}, suffers
from several drawbacks:\\
(i)At low densities $n\approx n_c$ and low temperatures, such as
explored in \cite{SKD_r(T)}, $T$ is of the order of
$\hbar/2\pi\tau $, and the requirement for the ballistic regime is
not satisfied (see, e.g.~\cite{das03}). In this case, one cannot
neglect the diffusive terms and crossover functions, as it was
done in Ref.~\cite{SKD_r(T)}.\\
(ii) In the vicinity of $n_c$, the temperature dependence of
$\sigma(T)$ is strongly affected by the procedure of sample
cooling \cite{cooldowns}, this regime is beyond the scope of
theory  \cite{ZNA}.\\
(iii) In the vicinity of $n=n_c$, the  ratio $\tau_q/\tau=3$, as
observed by SDK \cite{SKD_r(T),shashkin_SdH} is clearly anomalous;
this indicates that the standard models of disorder fail.\\
(iv) For low densities $n\sim 10^{11}$cm$^{-2}$ (and,
correspondingly, low values of $T_F^* \sim 2.5$K), the "metallic"
behavior corresponds to the temperature range where the valley
splitting $\Delta_v$ cannot be ignored. (Note that in our analysis
\cite{aleiner}, the valley splitting $\Delta_v \ll T$  and,
therefore, is insignificant). The sample-dependent parameter
$\Delta_v$ has not been measured for the SDK's sample, and SDK
made a doubtful assumption that the valley splitting $\gg 1K$
\cite{SKD_r(T)}, which contradicts their own Shubnikov-de Haas
data.  Correspondingly, in fitting the above Eq.~(3), the number
of triplet components was incorrectly reduced from 15 to
7. \\
For the aforementioned reasons, the interaction corrections theory
\cite{ZNA} can be hardly applied to the $\sigma(T)$-data analysis
at the densities $n\approx n_c$.  The ``accurate measurements in
the best samples'' in Ref.~\cite{SKD_r(T)} (though with mobility
$2.9\times10^4$cm$^{2}$/Vs \cite{SKD_r(T)} lower than $4\times
10^4$cm$^{2}$/Vs as in Fig.~2) have not been analyzed accurately
enough to provide a reliable quantitative information on the
renormalized effective mass. As the result of several unjustified
simplifications, the authors of Ref.~\cite{SKD_r(T)} overestimated
the increase of the effective mass with decreasing density.

\vspace{0.1in}

To summarize,
\begin{itemize}
\item Our $\rho(T)$ measurements performed over wide ranges of
densities and temperatures, being combined with the independent
measurements of renormalized $F_0^a(n)$ and $m^*(n)$ \cite{gm},
enable a rigorous test of the theory \cite{ZNA}. The theoretical
predictions are in a quantitative agreement with the experimental
data, in terms of either $\rho(T)$ or $\sigma(T)$, over the
density range where the theory can be applied ($\rho \ll h/e^2$,
$\Delta\rho(T)/\rho \ll 1$), and in the ballistic  $T$-range.
 \item The alternative attempt of such comparison
\cite{SKD_r(T)} in the regime  of low densities $n\approx n_c$ and
low temperatures suffers from several drawbacks, and does not
provide a reliable quantitative information on renormalization of
the effective mass.
\item We point out that the predictions of the
``screening'' theory of Ref.~\cite{GD} differ significantly from
those of the theory of interaction corrections \cite{ZNA} over a
wide range of densities, which is of primary importance for the
problem of ``metallic'' resistivity of high-mobility  Si MOSFETs.
\end{itemize}

\vspace{0.1in} Authors acknowledge support by the NSF, FWF
Austria, INTAS, RFBR, and the Russian Programs ``Integration of
high education and academic research'' and ``the State Support of
Leading Scientific Schools''.

\end{multicols}
\end{document}